# Avoiding another Green Elephant –
# A Proposal for the Next Generation HLA
# based on the Model Driven Architecture

*Dr. Andreas Tolk*
Virginia Modeling Analysis & Simulation Center (VMASC)
College of Engineering and Technology
Old Dominion University
7000 College Drive
Suffolk, VA 23435
(757) 686 - 6203
atolk@odu.edu

**ABSTRACT:** *When looking through the proceedings of the recent Simulation Interoperability Workshops, a lot of papers - some of them even awarded by the committee - are dealing with alternative concepts outside or beyond the High Level Architecture (HLA): Web Services, the extensible Markup Language (XML), Java Beans, Simple Object Access Protocol (SOAP), etc. Similarly, requirements driven by interoperability issues have resulted in the need to use meta modeling, adaptive models, and common repositories. The use of the Unified Modeling Language (UML) as a model description language is also rapidly becoming a standard. All these concepts have relations to the HLA, but they are not part of it. There seems to be the danger that HLA is overrun by respective developments of the free market and will become irrelevant finally.*

*A few years ago, another "Green Elephant" faced the same fate: The program language Ada was mandated to be used for military applications but became irrelevant for commercial applications. Underestimating the power of the free market - especially C++, Java, and the Internet protocols - Ada barely survived the wave of emerging technologies and tools. Today we might soon observe HLA fading into a similar insignificance if it doesn't continue to evolve.*

*The good news is that a potential solution already is formulated. First, the shortcomings of HLA are known and respective methods and processes are already identified in the papers mentioned before. Second, only recently the Object Management Group (OMG) introduced the Model-Driven Architecture (MDA) initiave as an approach to system-specification and interoperability based on the use of formal models. The core of the MDA concepts comprises the UML, XML as well as the related XML Metadata Interchange (XMI) specification, SOAP and other OMG standards. The concepts are designed to embrace OMG middleware solutions, i.e., the Common Object Request Broker Architecture (CORBA), as well as alternatives like Microsoft's Distributed Component Object Model (DCOM), Sun's Enterprise JavaBeans (EJB) etc. In other words, the solutions being perceived outside the HLA are in fact part of MDA. Thus, in order to keep the HLA relevant, where feasible, the HLA and MDA must merge into a new interoperability solution that is not only accepted by the SISO community, but one that also benefits from the overall efforts of the OMG.*

*This paper introduces the MDA concept and shows, how the HLA can be integrated to become a standard stub for simulation applications of legacy systems, systems under development, and systems of the future.*

# 1  Introduction

The program language Ada was one of the most modern and advanced of its time. It enabled and harmonized the concepts of the object oriented modeling and programming paradigm introducing a new quality to software development. Encapsulation, modules, components and patterns are only a few of the concepts that could have laid a solid ground for Ada to become the main programming language of the new object oriented paradigm. Consequently, military application designers with very demanding requirements became aware of the possibilities of Ada. A new star could have been born. Ada was introduced as a "standard" to the military. However, today only a small fraction of military software is written in Ada. Although still being very valuable in the academic world, the unbeaten champions are C++ and Java. What happened?

The reason for the tremendous distribution of C++ and Java lays not in the fact that these languages are superior to Ada. From a computer science point of view, these three languages are very similar and algorithms formulated in one of them can relatively easily be converted into another generally, in many cases even tool supported or automatically. It is also worth mentioning that Ada was – and still is – applied successfully in specific mission critical systems to implement respective firmware as well as in the more general domain of VHDL[1]. Why did C++ and Java take over the leading role as computer languages for software development and software engineering?

In summary, Ada did not fade due to the missing of application appropriateness. The reason for the success of the competing languages must lie elsewhere.

The author is convinced that the industrial support of C++ and Java, as well as the introduction of the Internet, are belonging to the main reasons for this development. Many technical solutions of Ada are still superior to C++ or Java, but while there grew up a tremendous support and tool industry for C++ and Java development, there was only very limited support for Ada. These effects were additionally multiplied by respective mandates forcing the rigorous application of military standards not only in military specific domains, but also in software domains for which the commercial sector already established efficient cost driven solutions. Together, this let do reinvention instead of reuse.

Overall, it was a misperception of the power of the free market that led to the decision to make Ada the first and only choice for military applications. The lack of commercial support in form of respective development and test tools, which were available to the C++ and Java community, forced the Ada community to live with inappropriate and costly implementation aids. These inconvenient conditions led to the term "The Green Elephant".[2]

Today, we have the mandate to use the High Level Architecture (HLA) to build military federations. The concept of HLA truly goes far beyond former approaches like the Distributed Interactive Simulation protocol (DIS) as well as the Aggregated Level Simulation Protocol (ALSP). The HLA introduced a new quality to the federation process, especially with respect to the technical aspect and taking processes and concepts into account like

- the Federation Development and Execution Process (FEDEP),
- the Functional Description of the Mission Space (FDMS) and
- Data Standards (DS) agreed to within the community.

However, again the doom of becoming another "Green Elephant" is perceivable at the horizon. Although the market for Modeling and Simulation (M&S) is everything but restricted to the military domain, attempts to bring the HLA to other applications domains in general is not a success story. There are some very interesting non-military applications that do use the HLA, but it hasn't become the backbone of simulation systems it could possibly be. Again, as with Ada, how did this happen?

As before, other powerful solutions out there support domains with very similar requirements and receive more industry support than HLA. The matured market of middleware as well as Internet based solutions has to be mentioned. Why should an industry that normally uses the Common Object Broker Request Architecture (CORBA) in their applications suddenly switch to the Run Time Infrastructure (RTI) as a new

---

[1]  **VHDL** = Very High Speed Integrated Circuits (VHSC) Hardware Description Language

communication standard? Why should industry leave their well-known CORBA-based IDL-tools to create an OMT-Model for information exchange?[3]

The challenge to couple systems using alternative standards is not new to the Simulation Interoperability Standards Organization (SISO) community. Especially when coupling simulation systems with real life systems – like weapon systems or Command and Control (C2) systems – this topic arises quite often [2]. Many of the new C2 systems are CORBA-based and additionally are using the Internet for information exchange. Furthermore, the modernized NATO AWACS and the Open Systems Avionics Technology being used for the F-16 and F-18 mid-life upgrade are CORBA-based. Overall, the number of CORBA-based solutions for military systems increased tremendously in all application domains of information technique.

In addition, respective industry standards already solved problems the HLA community has not been faced with so far. Furthermore, it almost goes without saying that a professional CORBA implementation based on the experiences of an open software industry is likely to be more efficient in its services than some RTI implementations. Adaptation and evolution of objects and systems are new fields to cope with as well. The management of large meta object groups and metadata interchange are directly derived from the experiences of the industry dealing with CORBA, which has lead to efficient and fast distributed new standards like the XML Metadata Interchange (XMI) format or the Meta Object Facility (MOF). The use of the Unified Modeling Language (UML) as a common standard for modeling is another unifying factor.

All these standards and their contribution to possible solutions to several problems with HLA have been presented in several excellent papers presented during recent Simulation Interoperability Workshops, e.g., [15, 16, 17]. The proposals to use these new solutions can be divided into two categories. The first set proposes the use of additional standards or respective derivations to improve the HLA following the motto "I want the best of both worlds." The second set uses the new standards instead of the HLA, following the motto "HLA is dead, long live CORBA/XML/…!" Both groups can be understood, but both groups have to consider the disadvantages that accompany their choices.

---

[3]   **RTI** = Runtime Interface;
      **IDL** = Interface Description Language;
      **OMT** = Object Model Template

## 2   Examples of Recent Approaches

It would go far beyond the scope of this paper to give a complete overview even of the requirements for new M&S approaches going beyond the limits of the HLA. Therefore, only a limited number of references to exemplifying papers will be given. Doubtlessly, many more valuable publications exist that cannot be referenced here. Therefore, this list is meant to be neither complete nor exclusive.

The following two papers presented during recent Simulation Interoperability Workshops show that HLA and CORBA are everything but exclusive concepts and may even complement each other:

- Reilly and Williams [7] showed how to use the CORBA middleware for sharing objects' data and methods over a network in a way consistent with the proposals of the HLA.

- Herzog et al. [8] proposed to use CORBA as a backbone for HLA solutions; a respective middleware that can be connected to an RTI that follows the standard as well as to an object request broker implementation.

In addition, Watson showed the potential of the use of open system solutions as proposed by the Object Management Group (OMG). With a respective presentation [13], he invited the M&S community to participate more actively in the commercial standardization process.

The idea to use CORBA within the HLA is everything but new. However, what may be new to many people is that an OMG Distributed Simulation Systems Specification - that maps the HLA to CORBA/IDL in a standardized manner [5] - exists. Although DMSO has sponsored this effort, the results have never been reflected in the appropriate manner by the respective SISO panels.

Nevertheless, CORBA is not the only technique being proposed to improve M&S in general and especially the HLA. Housand and Hudgins showed the feasibility to merge the Enterprise Java Beans (EJB) technology and the HLA concept to extend web server functionality in support of browser-based federation management [3].

Even within the series of the more or less HLA oriented Simulation Interoperability Workshops, some papers are trying to bring in new or additional ideas from outside of the M&S world. Some examples are:

- The necessity for proper documentation of federates and their behavior has been the topic of several papers. One way to deal with this, in a standardized manner, is the use of respective standards like the Unified Modeling Language (UML) as proposed in [16]. Many discussions in the respective Simulation Interoperability Workshop forums have proposed the idea to even replace the Object Model Template (OMT) standard with several possible alternatives (see also [1]); i.e., UML, XML and IDL.

- When standardizing the HLA via the IEEE, one of the requirements was to replace the first used BNF[4] for the OMT Data Interchange Format (DIF) with the new XML standard. In the meantime, a lot of other XML applications found their way into the Simulation Interoperability Standards Organization. Stytz and Banks having been among the strongest proponents within the M&S community. A good overview is given in [15].

- Another example of recent proposals is the paper given by Gustavson et al. [17] presenting the successful application of XML and the Simple Object Access Protocol (SOAP); another open commercially accepted and often applied standard for M&S integration. Again, the ideas of the HLA, especially its processes for federation development and execution (FEDEP), are incorporated.

- Similar observations can be made when evaluating the article on a web-based environment by Moradi, Svensson, and Ulriksson [19]. In their paper, they present a way to adapt HLA federates to the web in order to achieve distributed, component-based, and platform independent simulations. The underlying project on web-based HLA federations and simulations is sponsored by the Swedish Defense Research Agency (FOI).

Other workshops and symposia dealing with M&S issues are working on new alternatives parallel to the HLA as well. As many applications – among them also Command, Control, Communications, Computers, Intelligence, Surveillance, and Reconnaissance (C4ISR) systems – are using the Internet increasingly as their main communication backbone, among these alternatives web-based M&S applications are playing a special and important role:

- The Society for Computer Simulation International (SCS) has organized several workshops explicitly dealing with web-based simulation.

- The Modeling, Virtual Environments and Simulation (MOVES) Institute of the Naval Postgraduate School in Monterey, California, has launched several projects dealing with web-based simulation.

- The Department of Computer Science and Command, Control, Communications, and Intelligence (C3I) Center of the George Mason University is also supporting web-based and CORBA-oriented distributed virtual simulations.

- The Army's Joint Virtual Battlespace program has and continues to evaluate web-based approaches to federate the respective systems.

- The U.S. Joint Forces Command (JFCOM) continues to evaluate web-based tools and federates to increase the efficiency of large distributed computer-aided command post/field exercises supporting the training events of the Joint Warfighting Center (JWFC), as well as the experimentation events of the Joint Futures Laboratory (JFL).

Again, all these efforts are not limited to the HLA, but may use it when appropriate. However, the idea of using commercially supported standards like CORBA, UML, XML, SOAP, DCOM, and EJB etc. is central to all of them. Therefore, a merging of technology supporting these ideas would be the best solution. The Model Driven Architecture (MDA) may be the solution of choice, and the domain of M&S is still an open issue in so far published proposals.

# 3 The Model Driven Architecture $^{TM}$

To understand the potential of the MDA initiative, it is necessary to have a look at the executing organization first, the OMG. A short introduction to this institution will be followed by a technical overview of the MDA and a description of how to apply it. In addition, some short definitions and explanations of various techniques are given in the glossary at the end of this paper.

## 3.1 The Object Management Group

Eleven companies founded the OMG in April 1989. In October 1989, the OMG began independent operations as a not-for-profit corporation. Through the OMG's commitment to developing technically excellent, commercially viable and vendor independent specifications for the software industry, the consortium now includes over 800 members. The OMG is headquartered in Framingham, MA, USA and has international marketing offices in various countries all over the world along with a government representative in Washington, D.C.

---

4    **BNF** = Backus Naur Form

The OMG was initially formed to create a component-based software marketplace by supporting the introduction of standardized object software. The organization's charter includes the establishment of industry guidelines and detailed object management specifications to provide a common framework for application development. Conformance to these specifications makes it possible to develop a heterogeneous computing environment across all major hardware platforms and operating systems. Today, implementations of OMG specifications can be found on many operating systems across the world. OMG's series of specifications detail the necessary standard interfaces for Distributed Object Computing.

The OMG has led the way in providing vendor and language independent interoperability standards to the enterprise. Its goal is to enable a global information appliance. To this end, the infrastructure standard CORBA and the modeling standard UML have been introduced by the OMG. In addition to this, a very well accepted standardization process has been established, which as been improved over the recent years to develop – as well as refine - CORBA and UML. The Model Driven Architecture is the next step.

## 3.2 A Technical Overview of the MDA

The kernel ideas of meta modeling leading to the MDA initiave are not new to the Simulation Interoperability Standards Organization.

In various recent articles, the shortcomings of the Object Model Template (OMT) in comparison to data models using the IDEF1X standard or object models based on the UML already are described. In one of these papers, the idea of an M&S repository using meta models is introduced [1]. By the introduction of respective metadata, the shortcomings can be overcome without having to change the underlying standards. Figure 1 describes the idea of using an Information Resource Dictionary System (IRDS) to build up a common repository.

As the IRDS is for data, the Meta-Object Facility (MOF) is a general framework for modeling standards. Like HLA-OMT, IDEF1X and UML can be used to look at data to be shared in different ways – but looking at the same concept – a very similar view can be applied when looking at the different meta-levels of OMG artifacts (shared applications; e.g., programs, processes, objects).

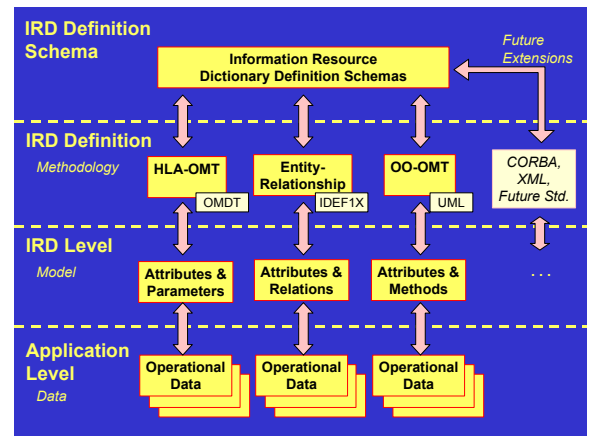

**Figure 1:** Applying the IRDS [1]

Bézivin explains this in [11]: As depicted in figure 2, UML describes the artifacts of Object-Oriented software systems on the level M2. Other meta models on the same level of abstraction may address data warehouses, organization, management, etc. Examples for such meta models are the Common Warehouse Metamodel (CWM) or the Unified Process Model (UPM). Emerging from the recognition that UML, CWM, UPM, etc. are only different meta-models in the software development landscape that can be brought together by introducing a common meta-meta model on the next level, the MOF was defined to fulfill these needs and requirements on the next level of abstraction.

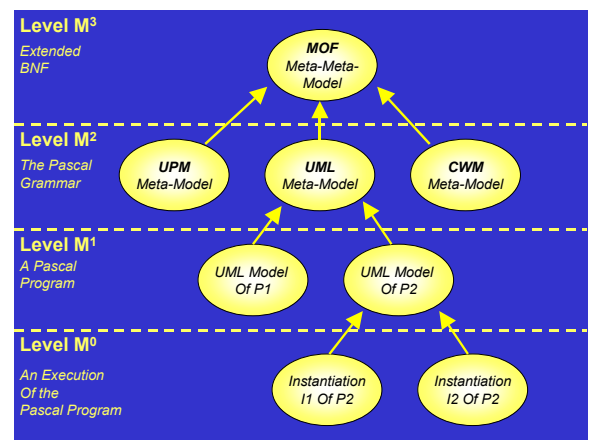

**Figure 2:** OMG four Layers Standard Modeling Stack

The respective OMG standard modeling stack comprises four layers in order to be able to cope with different levels of meta-modeling and abstraction.

The MDA is based on this idea of meta-modeling as well. It merges the different OMG standards having been developed and used separately so far into a common view by applying common meta models to them. However, it is not necessary to step in too deep into the meta worlds of modeling to understand the underlying concepts. The White Paper of Soley et al. [4] can be read and understood without requiring outstanding expertise in this domain.

The kernel idea is to use a common stable model, which is language-, vendor- and middleware-neutral. This model must be a meta-model of the concept. The MDA offers concepts for such a model. With such a model in the center, users having adopted the MDA gain the ability to derive code for various sub-levels. Even if the underlying infrastructure shifts over time, the meta-model remains stable and can be ported to various middleware implementations as well as platforms etc. Figure 3 shows the top-level view of the MDA comprising the stable model in the kernel.

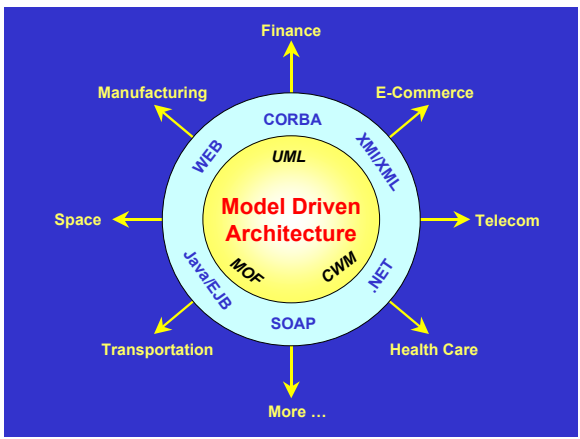

**Figure 3:** OMG Model Driven Architecture

As described in [4], the core of the architecture is based on OMG's modeling standards:

- Unified Modeling Language (UML)
- Meta-Object Facility (MOF)
- Common Warehouse Meta-model (CWM)

The different views of the core model will represent Enterprise Computing with its component structure and transactional interaction; another view will represent Real-Time computing with its special needs for resource control, etc. In any case, they will be independent of any middleware platform.

In the following subsection, the different phases leading from the model to the application, as described in detail in [14], will be shown. The various models and necessary profiles described are defined and exemplified in [10].

### 3.3 Applying the MDA

As already mentioned in the introduction, the MDA defines an approach to system specifications that separates the specification of the system functionality from the specification of the platform specific implementation. This is done by specifying standards to model the system in a reusable way. This allows two main applications:

- A system can be defined platform independently and then can be realized on multiple platforms through auxiliary mapping standards.

- Different applications can be integrated by explicitly relating their models, even if they do not run on the same platform type.

The first step when creating an MDA-based application is to create a Platform-Independent application Model (PIM). In the MDA, a model is defined to be a representation of a part of the function, structure and/or behavior of a system; i.e., the definition is usable in the M&S domain quite well. The PIM will be expressed in UML in terms of the appropriate core model. The core models are available in form of UML Profiles of which a number already are well along their way to be standardized by the OMG.

The next step – if the model shall run as an application – is to convert this model from general application to a Platform Specific Model (PSM). The PSM is derived from the PIM using standardized transformation rules. While the PIM defines the necessary functionality, the PSM specifies how this functionality is realized on a special platform.

This separation between requirement or general operationally driven activities and system specific functions, delivering the specification of these activities in a real implementation, is well known in the military domain. It reflects exactly the way to define C4ISR[5] architectures as it is stated in the respective architecture framework documents like introduced to

---

[5]  **C4ISR** = Command, Control, Communications, Computing, Intelligence, Surveillance, Reconnaissance

Simulations Interoperability Standards Organization in [12].

- The PIM is the equivalent to the system independent **Operational View** of the model ("What do we build? What functions are needed from the operational point of view?")

- The PSM can be interpreted as the **Systems' View** of the architecture of the model to be built ("How do we implement it?").

- The **Technical View** ("What standards do we need for the implementation?") also is reflected in the MDA by introducing respective vignettes in form of standard specific UML models – so called Stereotypes – in a latter step of the refinement process.

The last step is to generate code from these specific UML models. Figure 4 shows the flowchart for the overall process. In this figure, two sets of boxes are shown that haven't been dealt with explicitly so far, the Pervasive Services Model and the Domain Facilities Model:

- The **Domain Facilities Models** are directly connected to the CORBA domains standardized by the Domain Task Forces (DTF) of OMG members. Each DTF produces standard frameworks for standard functions in its application space. In [2] several examples are already given (including the reference to a DTF dealing with C4ISR issues from the CORBA standpoint).[6] In figure 3, the domains are shown as application domains like finance, space, telecom, etc. Distributed simulation systems as defined in [5] are among the candidates.

- The **Pervasive Services Model** comprises the definition of the set of essential services that are implemented as CORBA Services and Facilities within the CORBA environment; i.e., services like event notification, object security, transactions, etc. In addition, hardware and software attributes like scalability, real-time, fault tolerance, etc. may be modeled as well, if the user feels the need to do so in order to standardize the model.

---

[6] *The C4I DTF is a Domain Task Force of the Object Management Group (OMG) that operates under the Domain Technical Committee and is focused on systems that support crisis response, Search and Rescue (SAR), and military operations. The task comprises the definition of objects and services for Command, Control, Communications, Computers, Intelligence, Surveillance and Reconnaissance (C4ISR) systems together with consultation systems and sustainment disciplines (such as logistics, weather, air traffic control, etc.). Additional information is available at their web page: http://www.omg.org/homepages/c4i*

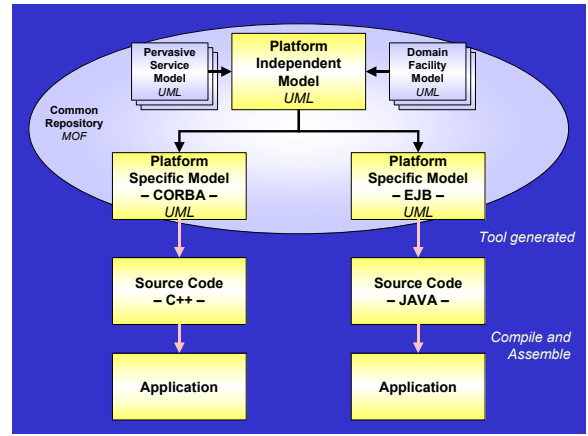

**Figure 4:** Application Generation using the MDA

Another aspect has to be mentioned as well. After having generated the PSM in UML, the generation of source code, configuration files, respective DTD[7] or XML schemas, SOAP, etc. - as well as the follow on compilation and assembling - can be supported – and in the most cases even be executed – by software tools.

When looking at the dates of the technical references to the MDA, it becomes obvious that this still is an evolving concept. Consequently, a lot of work still has to be done, but the potential is already obvious.

One of the domains that will be improved in the near future is to define and implement respective software tools to support the developers of MDA based applications. Actually, only a very limited number of MDA tools are available, but looking at the history of CORBA and UML, this is very likely to change in the near term. These tools will not only support the user in creating the PIM based on existing domain facilities and pervasive services, but also the common repository will hold a lot of ready to be reused solutions that only need to be modified or refined. Tools will help to map the PIM automatically to standardized PSM for most used platforms (like CORBA and .NET). They will also support the model mediation and harmonization of legacy models.

---

[7] **DTD** = Document Type Definition (used in XML to define the tags used in the documents)

However, the reengineering of legacy models and the harmonization of MDA based models is an aspect of its own. As most technical standards, the MDA facilitates respective efforts, but they do not come with the introduction of the MDA automatically. Additional management effort in form of alignment of participating processes is necessary. Object and data models that are badly aligned will not be harmonized by using the same standard to do the structuring. Respective methods to cope with this challenge of interoperability – like proposed in [6] and [9] – are not part of the MDA, although they are necessary steps on the way to interoperability. However, the MDA is an excellent choice to align architectures, or at least to describe architectures in a common language to facilitate the comparison and identification of possible misalignments.

At the end of this section it should be mentioned that the MDA has been named as a key trend in the software industry by PricewaterhouseCoopers in their recently published Technology Forecast for 2002-2004 [18], which reports that the MDA is poised to revolutionize the software design and development process. It is therefore very likely that the MDA will become a success story like the middleware solution CORBA as well as the modeling language UML already are. The author is convinced that the MDA approach will help the commercial industry to reach a new level of interoperability within this decade.

# 4 Merging the Concepts and Ideas of the HLA into the MDA

Although there have been some hints in the former sections, it may not be clear why the MDA influences the future of M&S in general and specifically the future of HLA. Therefore, the kernel activities perceived to be necessary on the short and mid term are summarized in this section.

It should be stated very clearly that the author believes that the HLA has been a success story so far. The HLA codifies and standardizes a set of simulation services that have existed before as stove pipe or proprietary solutions, and that are essential to simulation systems in any case. One has to distinguish between the principles of the HLA and implementing concepts. The author is convinced that the principles of the HLA are outstanding and should find their way into a much broader community.

However, the implementation of these principles reveals much room for improvement. While the HLA can help the MDA to improve in its principles concerning distributed simulation systems, the MDA can help the HLA implementers to improve their efforts based on the experiences of the OMG partners and the related software industry.

## 4.1 Why should we merge?

The need for incorporation of open systems solutions has been discussed quite often recently, especially when addressing integration issues with real systems (e.g., see [2]). Partners from the commercial sector who build, e.g., telecom systems or flight coordination systems, will very likely use respective standards like the MDA as they are now using CORBA. In addition, the MDA is not in contradiction with major standards used in the military domain. CORBA already is in the U.S. Department of Defense Common Operating Environment (COE) and the MDA will find is way into it as well.

Additionally, avionics as well as vectronics systems are more and more CORBA-based and will use the MDA as well. The use of their PIM and their PSM will facilitate the construction of simulators, as it will be much easier to understand and model the system starting with a standardized description of it. Also, the integration of these systems as federates into M&S federations or the design of alternative interfaces to couple them will be supported by the standardized meta models used within the MDA.

Furthermore, overarching new concepts like the Future Combat Systems (FCS) that need to integrate numerous systems and concepts – from the sensors vs. the weapon systems to the command and control "System of systems" – are also on the list of potential MDA users. Another candidate, who is actually using CORBA as its communications backbone, is the Test and Training Enabling Architecture (TENA), which is designed to enable interoperability between live test and training ranges, respective facilities and simulation systems of all application domains. A short description of TENA and further references are given in [20]. Even if these system families – FCS and TENA – decide to stay with CORBA and will not switch to the MDA immediately, the explicit use of CORBA as a kernel technology in the MDA demands them to deal with this new approach (and the existence of a CORBA PIM will facilitate the development of an FCS PIM or a TENA PIM anyhow).

The same statements are true for the XML standards family as well. In other words, all systems using XML interfaces are potential candidates for MDA applications, as there will be standardized ways in the near future to migrate XML based solutions to the MDA. This is especially true for XMI based solutions.

Every simulation or federate that wants to exchange data with such MDA based systems has to know the underlying concepts. When embedding the MDA system into a federate – or vice versa – it becomes obligatory to know and understand the architectural concepts. The use of aligned architectures has already been recommended more than once. The MDA enables even more to use the same standardized kernel to build models on that really have the potential to be interoperable from the definition and design phase on.

### 4.2 How should we merge?

There are many ways to participate in the MDA efforts. Therefore, the following proposals are neither complete nor exclusive and discussions on these and additional points are encouraged and welcome:

- **Domain Facilities**: As already proposed in [2], the standardized domain facilities of related domains, e.g. the C4ISR domain, should be incorporated into the HLA. The use of this standard should be mandatory. Additionally, the HLA community should actively participate in the evolution of the Distributed Simulation Systems Domain [5] making this facility the M&S interoperability standard of the next generation.

- **Pervasive Services**: The services provided by the RTI, as defined by the HLA, have to be harmonized with the pervasive services (see [8] for how this can be done). In addition, services needed for M&S – which have not yet been added as PSMs - have to be standardized and integrated. The works of the Naval Postgraduate School as well as the German efforts as described in [8] have to be mentioned explicitly trying to benefit from RTI services enriched by additional CORBA services. This is a very promising direction.

- **RTI as Middleware**: Instead of the actual solution, in the future a standardized PSM of the MDA based on the general PIM for Distributed Simulation Systems should define the RTI. This will place the RTI side-by-side with middleware solutions like CORBA, EJB, etc. In addition to this standard-oriented work, the implementation of tools as mentioned in subsection 3.3 will facilitate the implementation of simulation applications.

- **Federation Development Tools**: Many recent presentations introduced tools facilitating the development and management of federations. Especially tools connecting the interface of federates to the RTI and vice versa as well as tools encapsulating the RTI in a more convenient form were often presented. The MDA offers a possibility to standardize these efforts by introducing overarching PIM. Instead of building new tools, industry could focus on more efficient automatic mapping tools avoiding the reinvention of the wheel with every new middleware approach.

- **Data Engineering**: The MDA doesn't solve the problems data engineering is dealing with. Data engineering comprises the tasks of

  - *Data Administration:* What data is available in what format in which sources and is accessible using what type of media, etc.?

  - *Data Management:* What is the semantic of respective data? What standardized data elements can be used to be mapped to the data to be standardized?

  - *Data Alignment*: How well do the data of the source match the data of the sink? Which data elements are missing?

  The methodologies of data engineering that have been presented in [6] and [9] must be adapted to meet the requirements of the MDA, and then they can be used for efficient reengineering of models into the MDA.

To summarize, many of the insights presented at respective Simulation Interoperability Workshops can be brought to a broader community by lifting them to the next abstraction level of meta modeling. E.g., the harmonization of information exchange requirements doesn't come with introducing new data standards automatically. To standardize the semantics in form of a common ontology is still necessary. Neither will the processes of setting up a federation be harmonized by introducing the MDA. The M&S community already has experiences in these respective problem domains and can contribute to reach efficient interoperable solutions be adapting their solutions to the MDA, which would be of benefit for all sides.

## 5    Conclusions and Recommendations

Although the respective efforts are still very young, the MDA approach is building on solid ground using and merging with established and matured standards like UML, CORBA, XMI/XML, EJB, and CWM etc. In addition to the already mentioned advantages of the OMG, many of these standards are not only supported by participating vendors, but they are supported by the broad open source community as well, which will add tremendous resources of professional solution developers, implementers, and peer reviewers to the overall process.

In summary, the MDA train just left the station and is gaining speed, and we can either jump onto the train and join the ride, or we may stay were we are and be overrun by it.

The Simulation Interoperability Standards Organization is the standardizing facility for M&S, although mainly in the military domain. It can play an important role in the process of integrating the HLA into the MDA making it the commercially accepted and supported way to integrate M&S systems. The principles of the HLA are matured enough to improve the MDA in the domain of distributed simulation. On the other hand, the OMG can help to "revitalize" the HLA and improve the respective implementations. Therefore, both sides would benefit from a closer relation.

To do this, liaisons have to be installed and refreshed and the need to participate in the MDA activities has to be made well known within the community via these liaisons.

The HLA has been proven to be an efficient platform for M&S integration and delivered a new level of interoperability. Therefore, the M&S community should participate in the overarching MDA efforts by integrating the HLA tools for federation management and federation development as well as by integrating MDA-based RTI implementations. These efforts will help us to reuse our lessons learned after the paradigm shift to use meta modeling as a new system development and integration methodology. In addition, these efforts integrated into the MDA efforts will also help the M&S community to reach new users via the OMG who had no contact to M&S applications before.

It is urgently recommended to the Simulation Interoperability Standards Organization to participate in the MDA efforts by

- Establishing liaisons to the OMG
- Establishing a working group dealing with the MDA and the integration of the RTI, i.e.,
- Support the MDA with respective HLA tools

Needless to say, that the Simulation Interoperability Standards Organization can be a tremendous facilitator and supporter in the overarching process by providing the platform and forums for these works and efforts to be done to insure interoperability for simulation applications of legacy systems, systems under development and systems of the future.

# 6    Glossary

*Within this short glossary, some of the specifications issued by the OMG will be highlighted for the reader who isn't too familiar with the software development community and respective standards. This list is only a small subset of alternatives of competitive as well as completing techniques:*

- *The **Object Management Architecture** (OMA) builds the backbone of applications based on the Common Object Request Broker Architecture (CORBA). It categorizes objects into the four categories services, facilities, domain objects, and application objects. The OMA abstracts out this common functionality from CORBA applications into a set of standard objects that perform standard, clearly defined functions, accessed through standardized interfaces.*

- *The **Unified Modeling Language** (UML) is a graphical language that expresses application requirements analysis and program design in a standard way. By a well-defined syntax and semantics it is a basis for a widely spread common model representation supported by a great number of software tools.*

- *The **Meta Object Facility** (MOF) provides a universal way to describe concepts, also referred to as meta models. It is the highest level of abstraction being standardized by the OMG so far. MOF is utilizing UML as the modeling language for concepts and meta models.*

- *The **Extensible Markup Language** (XML) is designed to improve the functionality of the Web by providing more flexible and adaptable information identification. It is a standardized file format on the basis of the Standard Generalized Markup Language (SGML) and became a wide spread standard for web applications within the last couple of months.*

- *The **XML Metadata Interchange** (XMI) is a standard for representing, sharing, and interchanging data and meta data on the basis of*

XML. It is designed to be machine readable as well as human readable and builds in parallel a basis to structure a common repository. XMI is a stream format for interchange of metadata including the UML models created during analysis and design activities, thus XMI bridges the gap between XML and Objects.

- The **Common Warehouse Metamodel** (CWM) standardizes a basis for data modeling commonality within an enterprise, across databases and data stores. Building on a foundation meta model, it adds meta models for relational, record, and multidimensional data; transformations, OLAP, and data mining; and warehouse functions including process and operation. CWM maps to existing schemas, supporting automated schema generation and database loading.

- The **Universal Process Model** (UPM) provides a high level overview. It describes the basic process steps and provides general guidance on their role and order.

- The **Simple Object Access Protocol** (SOAP) is a protocol for exchange of information in a decentralized, distributed environment. It is an XML based protocol.

- The **Java Metadata Interface** (JMI) is an upcoming standard to provide standard Java interfaces for software models.

- The **Enterprise Distributed Object Computing** (EDOC) Profile includes several models for application oriented software structures, among others for Java, Enterprise JavaBeans (EJB), Flow Composition Models (FCM), and others.

- The **Enterprise Application Integration** (EAI) specification defines models for text based messages, C, C++, COBOL, and more. By providing mappings of data types between heterogeneous federations based on mixed languages, it facilitates the development of respective enterprise applications.

- The **Model Driven Architecture** (MDA) is the "melting pot" for all these different methods and methodologies. It is the logical next step to expand the OMA to use the broadened possibilities in a coordinated manner by consequently using standard driven meta-models to find common kernels as a basis for common models – and therefore common understanding of the nature of the problem of federated solutions.

In addition, the homepage of the Object Management Group was a very valuable source for the OMG-papers as well as related documentation: http://www.omg.org

## Acknowledgement

*The author thanks all individual professionals that read and commented on this paper helping to improve it from a first idea into this actual proposal. Among others, the following friends participated in the process predominantly: Michael R. Hieb (IITRI/ABTech), Mark A. Phillips (VMASC/ODU), and Donald H. Timian (Northrop Grumman).*

## Authors' Biography

**ANDREAS TOLK** is Senior Research Scientist at the Virginia Modeling Analysis and Simulation Center (VMASC) of the Old Dominion University (ODU) of Norfolk, Virginia. He has over ten years of international experience in the field of Applied Military Operations Research and Modeling and Simulation of and for Command and Control Systems. In addition to his research work he gives lectures in the Modeling and Simulation program of ODU. He also gives lectures for NATO as a Subject Matter Expert on Military Decision Making.

*The paper has also been published on the OMG website comprising presentations and discussion papers dealing with the Model Driven Architecture – see http://www.omg.org/mda/presentations.htm*